\documentclass{iopart}
\usepackage{cite}

\jl{6}        
\eqnobysec    

\def\beq{\begin{equation}}
\def\eeq{\end{equation}}
\def\rmd{{\rm d}}

\begin{document}

\title[Petrov classification of perturbed spacetimes: the Kasner example]
{Petrov classification of perturbed spacetimes: the Kasner example}

\author{
Christian Cherubini${}^{\ddag}{}^{\ast}{}^{\S }$,
Donato Bini${}^{\dag\,\ast}$,
Marco Bruni${}^{\S }$,
Zoltan Perjes${}^{\P}$}

\address{
  ${}^{\ddag}$\
 Faculty of Engineering, University Campus Bio-Medico of Rome, 
via E. Longoni 47, 00155 Rome, Italy.}

\address{
  ${}^{\ast}$\
International Center for Relativistic Astrophysics, I.C.R.A., 
University of Rome ``La Sapienza,'', I--00185 Rome, Italy.}

\address{
  ${}^{\S}$\
Institute of Cosmology and Gravitation, University of Portsmouth,
Portsmouth, PO1 2EG, UK.
}

\address{
  ${}^{\dag}$\
Istituto per le Applicazioni del Calcolo \lq\lq M. Picone\rq\rq, C.N.R.,
   I-- 00161 Roma, Italy.}

\address{
  ${}^{\P}$\
KFKI Research Institute for Particle and Nuclear Physics, H-1525, Budapest 114, P.O.B. 49, Hungary.
}

\begin{abstract}
In this paper we consider vacuum Kasner spacetimes, focusing on those that  can be parametrized as  linear perturbations of the special  Petrov  type $D$ case. For these quasi-$D$ Kasner models  we first investigate  the modification to the principal null directions, then 
a Teukolsky Master Equation for fields of any spin, considering in particular the quasi-$D$ models as curvature perturbations of the type $D$ background. 
Considering the speciality index and the principal null directions and comparing the results for the exact solutions and those for the perturbative ones,  this simple Kasner example allows us 
to clarify that perturbed spacetime  do not retain in general the speciality character of the background. There are four distinct principal null directions, although they are not necessarily first order perturbations of the  background principal null directions, as our example of the quasi-$D$ Kasner models shows. 
  For the quasi-$D$ Kasner models the use of a Teukolsky Master Equation, a classical tool for studying  black hole perturbations, allows us to show, from a completely new point of view, the well known absence of gravitational waves in  Kasner spacetimes. This result, used together with an explicit expression of the electric Weyl tensor in terms of Weyl scalars, provides an example of the fact that the presence of transverse curvature terms does not necessarily imply the presence of gravitational waves.

\end{abstract}

\section{Introduction}

In this paper we apply, to the vacuum Kasner spacetime models \cite{DIN}, 
some mathematical tools developed in the context of black hole physics,
{\it i.e.} the speciality index (SI) 
\cite{BC} and the Teukolsky Master Equation (TME) \cite{Teukolsky2}. 
The first one indicates whether a given spacetime is algebraically special and is used here, in conjunction with a study of the principal null directions (PNDs), to discuss
how linear perturbations may change the speciality condition. Kasner models are Petrov type $I$ (general), yet they are simple enough that a direct comparison between the exact and perturbative solutions is possible, which is very useful to clarify, in a transparent way, some results of perturbation theory. 
In particular, the usually defined SI $\mathcal{S}$ only changes at second order for perturbations of a generic type $D$ background \cite{BC}, yet one would expect the perturbed spacetime to be of general   type $I$ Petrov type. Considering quasi-$D$ Kasner models that can be parametrized  as linear perturbations of the special  Petrov  type $D$ case, we show that they have four distinct PND, as expected and in complete agreement with the exact models. However, the PND of a linearly perturbed model are not necessarily first order perturbations of the  background PND, as we show for our quasi-$D$ Kasner models.

The TME is a single partial differential equation 
describing the behavior of perturbations  of a given Petrov type $D$ background due
to fields of any spin $(0,\pm 1/2,\pm 1,\pm 3/2, \pm 2)$; it is therefore a classical tool for studying perturbations of black holes, in particular gravitational radiation \cite{Teukolsky2}. It  is used here to show the absence of gravitational waves in the quasi-$D$ Kasner models. This confirms,  with  a completely new approach, the well known absence of gravitational radiation in general type $I$ Kasner spacetimes, 
 shown for example by the vanishing in Kasner models of the gravitational superenergy flux  (see {\it e.g.} \cite{BV}). Then, we use the non vanishing of the  Weyl scalar $\psi_0$ and $\psi_4$ in a general Petrov type $I$ Kasner model (either an exact or pertubative solution)  to illustrate that the non-vanishing of these transverse contributions to the electric Weyl tensor (the only curvature contribution in vacuum to the Jacobi or geodesic deviation equation) does not necessarily imply the presence of gravitation radiation. 

As the Newman-Penrose formalism will be adopted here, conventions and notation will follow
the existing literature on the subject~\cite{Teukolsky2,ChandraBook}, including that for the
metric signature $(+,-,-,-)$.

\section{Petrov classification: speciality index and principal null directions}

Defining the complex tensor 
$\tilde C_{abcd}=C_{abcd}-i{}^*C_{abcd}$, one can introduce the two
complex curvature invariants, in tensor and Newman-Penrose (NP) form
\begin{equation}
I=\frac 1{32}\tilde{C}_{abcd}\tilde{C}^{abcd}=(\psi _0\psi _4-4\psi _1\psi
_3+3\psi _2^2)
\end{equation}
and 
\begin{eqnarray}
\fl\quad
J&=&\frac 1{384}\tilde{C}_{abcd}\tilde{C}^{cd}{}_{mn}\tilde{C}^{mnab}
=(\psi_0\psi _2\psi _4-\psi _1^2\psi _4-\psi _0\psi _3^2+2\psi _1\psi _2\psi
_3-\psi _2^3)\,.
\end{eqnarray}
These can be used to define the speciality index~\cite{BC,beetle} 
\begin{equation}
\mathcal{S}=\frac{27J^2}{I^3}\,;  \label{SPECT}
\end{equation}
with its value this demarcates, in an invariant way,
the transition from   certain algebraically special solutions ($\mathcal{S}=1$) 
and the general Petrov type $I$ ($\mathcal{S}\neq 1$)~\cite{kraetal}. 
This quantity can be easily evaluated in the case of 
the vacuum Kasner~\cite{LL} metric: 
\begin{equation}
\rmd s^2=\rmd t^2-t^{2p_1}\rmd x^2-t^{2p_2}\rmd y^2-t^{2p_3}\rmd z^2\,,\quad  \label{LLLL}
\end{equation}
where 
\begin{equation}
p_1+p_2+p_3=p_1^2+p_2^2+p_3^2=1 \label{constr}
\end{equation}
 and for the moment there is no preferential ordering of the Kasner
indices. 
 
Let us now introduce a Newman-Penrose tetrad
\begin{eqnarray}  
l&=& \frac{1}{\sqrt{2}}[\partial_t+t^{-p1}\partial_x], \quad  n=\frac{1}{\sqrt{2}}[\partial_t-t^{-p1}\partial_x] \,, \label{tetrade} \\
m &=&\frac 1{\sqrt{2}}[t^{-p_2}\partial_y+it^{-p_3}\partial_z] \,, \label{tetradeb}
\end{eqnarray}
which gives the non zero spin coefficients: 
\begin{eqnarray}
\fl\quad
\mu &=&\frac 1{2\sqrt{2}t}(p_2+p_3)=-\rho,\quad \epsilon = \frac{p_1}{2\sqrt{2}t}=-\gamma,\quad \lambda =\frac 1{2\sqrt{2}t}(p_2-p_3)=-\sigma
\,, 
\end{eqnarray}
the non zero Weyl scalars: 
\begin{eqnarray}
\label{weylscal}
&&\psi _0=\psi_4=\frac{p_1(p_2-p_3)}{2t^2}, \quad \psi _2=-\frac{p_2p_3}{2t^2}\,, \
\end{eqnarray}
and generates the time independent SI 
\begin{equation}
\mathcal{S}=\frac{27}4p_3^2(1-p_3)\,.  \label{SPECTRX}
\end{equation}
This tetrad is usually called \lq\lq transverse"  because of the property $\psi _1=\psi_3=0$ \cite{beetle,Szekeres}, as it is explicit from equation (\ref{bella}) below (see also the discussion at the end of Sections 3 and 4). From (\ref{SPECTRX}) clearly the Petrov type is $I$ in general; the null tetrad  may be said in canonical form because $\psi_0=\psi_4$ \cite{Pollney}. 

From (\ref{SPECTRX}) $\mathcal{S}$  results well defined for any
Kasner solution, while in general this may not be the case~\cite{beetle}.
The Kasner metric admits two special subcases when two of the $p_i$ indices are equal: it then follows from (\ref{constr}) that either $p_1=p_2=0$, $p_3=1$ (and permutations) and the spacetime is flat in this case, or  $p_1=-1/3$, $p_2=p_3=2/3$ (and permutations) and one has the Kasner LRS type $D$ solution, with $\mathcal{S}=1$, with a spindle-like
singularity~\cite{kraetal,DIN}. As it was shown by Geroch~\cite{geroch, paiva},
this type $D$ model can be obtained as the limit for $M\to \infty$ of the Schwarzschild
solution, provided a specific coordinate transformation.
For the type $D$ case with $p_2=p_3=2/3$ the null tetrad above is a 
principal one, with $l$ and $n$ aligned along the PNDs of the Weyl tensor in this case, and is also canonical\cite{Pollney},  
with $\psi_0=\psi_4=0$. The same tetrad is not canonical for the other two physically equivalent type $D$ cases $p_1=p_2=2/3$ and $p_1=p_3=2/3$, with $\psi_0=\psi_4\neq0$.

The orthonormal frame naturally associated to (\ref{tetrade})-(\ref{tetradeb}) is 
\begin{eqnarray}\label{orto}
e_0&=&\frac{1}{\sqrt{2}}(l+n)=\partial_t=u\,,\, \quad  e_1=\frac{1}{\sqrt{2}}(l-n)=t^{-p_1}\partial_x\,,\nonumber \\
e_2&=&\frac{1}{\sqrt{2}}(m+\bar m)=t^{-p_2}\partial_y\,,\quad  e_3=\frac{1}{\sqrt{2}i}(m-\bar m)=t^{-p_3}\partial_z\,.
\end{eqnarray}
It is adapted to the static preferential observers with 4-velocity $u$ who use the Killing vectors $\partial_x, \partial_y, \partial_z$ to build their spatial axes and therefore directly observe the homogeneity of the spacetime. They also observe a purely electric Weyl tensor
\begin{eqnarray}
\label{elemagn}
E(u)&=&\frac{p_1p_3}{t^2}\, e_1\otimes e_1+ \frac{p_1p_2}{t^2}\, e_2\otimes e_2+ \frac{p_2p_3}{t^2}\, e_3\otimes e_3,\label{EW} \\
H(u)&=& 0,\label{MW}
\end{eqnarray}
the electric and magnetic part of the Weyl tensor being
$$
E(u)_{\alpha\beta}=C_{\alpha\mu\beta\nu}u^\mu u^\nu, \qquad H(u)_{\alpha\beta}=-{}^*C_{\alpha\mu\beta\nu}u^\mu u^\nu ,
$$
respectively.
It is also useful to give here the expression of $E(u)$ in terms of NP quantities in a transverse NP frame
\begin{equation}\label{bella}
\fl\qquad
  E(u)=\mathrm{Re}(\psi_2) e_{\mathrm{C}} -\frac{1}{2}\mathrm{Re}(\psi_0+\psi_4)e_{\mathrm{T}+} +\frac{1}{2} \mathrm{Im}(\psi_0-\psi_4)e_{\mathrm{T}\times}, 
\end{equation}
where $e_{\rm{C}}= e_1 \otimes e_1 +e_2\otimes e_2 -2e_3\otimes e_3$, 
$e_{\mathrm{T}\times} = e_1\otimes e_2 +e_2\otimes e_1 $ and 
$e_{\mathrm{T}+}= e_1\otimes e_1 -e_2\otimes e_2$ respectively represent a Coulombian and two transverse basis tensors.
This relation is further simplified if the transverse frame is also canonical (as the frame (\ref{tetrade})-(\ref{tetradeb}) used in this case), giving the result (\ref{elemagn}).

We now want to compare the above results for the Petrov classification of  the Kasner exact solutions  and results for Kasner models in a neighborhood $\varepsilon$  ($\varepsilon $ being a small quantity) of the type $D$ case, using $p_3=2/3+\varepsilon$. Guided by the results for the exact solutions, we will be able to interpret the perturbative results in a coherent framework.
Expanding the SI around the type $D$ case, one gets $\mathcal{S}=1-27/4\varepsilon ^2-27/4\varepsilon ^3
$, which confirms the perturbative result found in  \cite{BC} for a generic type $D$ background starting from the canonical tetrad for type $D$ $(\psi _0^{(0)}=\psi
_1^{(0)}=\psi _3^{(0)}=\psi _4^{(0)}\equiv 0)$, with the index $N=0,1,...$ in $\psi _n^{(N)}$
denoting the perturbative order:
\begin{equation}
\mathcal{S}=1-3\frac{\psi _4^{(1)}\psi _0^{(1)}}{(\psi _2^{(0)})^2}\varepsilon
^2+o(\varepsilon ^3)\,. \label{expS}
\end{equation}
From this result, solutions like this type $D$ Kasner one could seem to
change perturbatively their  Petrov type at
second order only. Moreover if the gravitational perturbations are
algebraically special~\cite{ChandraBook,chandra2} ({\it i.e.} either $\psi _0^{(1)}$
or $\psi _4^{(1)}$ vanishing) the change should start at third order. One is therefore tempted to conclude from (\ref{expS}) that a spacetime that is a linear perturbation of a type $D$ background retains the type $D$ character at first order \cite{zoltan}.
On the other hand, it seems very difficult to believe that linear perturbations of an 
algebraically special solution do not alter the speciality conditions, {\it i.e.} do not modify the 
number of the PNDs. 
In other words, the fact that $\mathcal{S}$ only changes at second order seems to have more to do with its definition that with a real speciality of the Petrov type of the linearly perturbed spacetime.
Indeed, one can redefine the speciality index using a monotonic function of $\mathcal{S}$ in order to have perturbative changes starting at first order in $\varepsilon$: this can be easily shown by using for instance  the normalized index $\mathcal{S}_{\rm norm}=\sqrt{|\mathcal{S}-1|}$. Thus the point is, that once first order dynamical variables such as the perturbed metric are chosen, derived quantities such as $\mathcal{S}$ and $\mathcal{S}_{\rm norm}$ may or may not be linear in the perturbations, a fact that  only reflects the way they are defined.

In order to completely clarify this issue of the Petrov classification of perturbed solutions, we are now going to build up, as a concrete example, the  principal null directions of the general Kasner vacuum models~(\ref{LLLL}) and to  expand them perturbatively around the  type $D$ case $p_2=p_3=2/3$. Of course, as soon as $p_2-p_3$ is nonzero (no matter how small) 
it is clear from  (\ref{weylscal}) and (\ref{SPECTRX}) that  the Petrov type of any Kasner solution 
is general, so that there exist four distinct PNDs. As a consequence, we will conclude in particular that
there must be four perturbed distinct PNDs for the quasi-$D$ Kasner  model, as then confirmed by our perturbative analysis.
We proceed as follows.
First of all, let us recall that, in general, once one has found a generic NP frame with its associated
Weyl scalars, given in our case by equations (\ref{tetrade})-(\ref{tetradeb}) and (\ref{weylscal}), 
the PNDs of that manifold are~\cite{ChandraBook} 
\begin{equation}
{h}_{(j)}^a\propto {l^a}+b_{(j)}^{*}{m^a}+b_{(j)}{\bar{m}^a}
+b_{(j)}b_{(j)}^{*}{n^a}\,,  \label{PNDEQ}
\end{equation}
where the $b_{(j)}$ are the four roots ($j=1,2,3,4$) of the equation (assuming $\psi_4\neq0$)
\begin{equation}
\psi _0+4b\psi _1+6b^2\psi _2+4b^3\psi _3+b^4\psi _4=0\,.
\end{equation}
In our case they result in 
\begin{equation}
b_{(j)}\equiv b_{(\zeta ,\eta )}=\zeta \sqrt{C+\eta \sqrt{C^2-1}}
\,,\qquad C=\frac{3p_2p_3}{p_1(p_2-p_3)}\,,  \label{radici}
\end{equation}
where $\zeta $ and $\eta $ assume independently the values $\pm 1$. Assuming, with no loss of generality, 
the following ordered parametrization for $p_1,p_2,p_3$~\cite{kraetal,LL} 
\begin{equation}
-\frac 13\le p_1\le 0,\quad 0\le p_2\le \frac 23,\quad \frac 23\le p_3\le
1\,,\qquad \Longrightarrow C\ge 3\,,  \label{ordered}
\end{equation}
the four roots (\ref{radici}) result real, simplifying considerably the
calculations. We stress that the PNDs (\ref{PNDEQ}) have still the freedom
to be rescaled by a  multiplicative factor, which we will fix in
order to  recover the correct  $l$ and $n$ in (\ref{tetrade}) in the type $D$ limit. 
To
this purpose let us study first the case $\eta =1$ (still maintaining the
freedom $\zeta =\pm 1$) in (\ref{radici}), denoting it as $b_{(+)}$. In
this case, the first two PNDs, using (\ref{PNDEQ}) and (\ref{tetrade})-(\ref{tetradeb}), 
result in 
\begin{equation}
{h}_{(+)}^a\equiv \frac{1}{b_{(+)}^2}\,\left[ {l^a}+b_{(+)}{m^a}+b_{(+)}{
\bar{m}^a}+b_{(+)}^2{n^a}\right] \,,  \label{prime}
\end{equation}
For $p_1=-1/3$, $p_2=p_3=2/3$ these directions give exactly the type $D$ limit of $n$ in equation (\ref
{tetrade}), which is a double principal null direction for the spacetime. Let us
study now the case $\eta =-1$ in (\ref{radici}), denoting it as $b_{(-)}$.
In this case the remaining two PNDs, using (\ref{PNDEQ}) and (\ref{tetrade})-(\ref{tetradeb}), 
are 
\begin{equation}
{h}_{(-)}^a\equiv \,{l^a}+b_{(-)}{m^a}+b_{(-)}{\bar{m}^a}+b_{(-)}^2{n^a}
\label{seconde}
\end{equation}
where we point out the absence of the rescaling with respect to the previous
PNDs. In the same limit as above, we obtain exactly the other repeated PND $l$ 
of the type $D$ case in (\ref{tetrade}). It can be easily verified that in general
these four PNDs are not linearly independent because the $z$ component is
always zero. This is expected because of the vanishing of the magnetic part
of the Weyl tensor (see ref. \cite{kraetal}, page 54). 

Note that although the $b_{(+)}$ roots in (\ref{radici}) blow up in the type $D$ limit, while the $b_{(-)}$ vanish, the PNDs above are always well defined. However, although the type $D$
limit of all the PNDs exists, they don't admit  a  Taylor expansion around this point. 
With no loss of generality, we demonstrate this property in the simpler case of the ordered
range (\ref{ordered}).
Assume again $p_3=2/3+\varepsilon$, with $\varepsilon>0$ to satisfy (\ref{ordered}). The perturbation of Kasner
constraints (\ref{constr}) leads to $p_1=-1/3+O(\varepsilon ^2)$ and $
p_2=2/3-\varepsilon +O(\varepsilon ^2)$. Although the null vectors $l,n,m$
of (\ref{tetrade})-(\ref{tetradeb}) have smooth Taylor expansions in $\varepsilon $, 
for the PNDs (\ref{prime}) and (\ref{seconde}) we obtain
\begin{eqnarray}
&&{h}_{(-)}^a=\frac 1{\sqrt{2}}\left[ 1+O(\varepsilon
),\,t^{\frac 13}+O(\varepsilon ),\,-\zeta t^{-\frac 23}\varepsilon ^{\frac
12}+O(\varepsilon ^{\frac 32}),\,0\right] \,, \\
&&{h}_{(+)}^a=\frac 1{\sqrt{2}}\left[ 1+O(\varepsilon
),\,-t^{\frac 13}+O(\varepsilon ),\,-\zeta t^{-\frac 23}\varepsilon ^{\frac 12}+O(\varepsilon
^{\frac 32}),\,0\right] \,,
\end{eqnarray}
clearly not a  Taylor series, but nonetheless 
manifesting the splitting of the PNDs in passing from the Petrov type $D$ to the type $I$.
This peculiar phenomenon regards the $y$ component of the null directions, while
the $z$ component of these PNDs remains always zero. 
Analogous results will be obtained using  another choice of order for the Kasner parameters.
The nonanalytic behaviour of the principal null directions in the parameter space 
is to be expected at the algebraically special loci. Small changes in the parameters
have the effect that the coincident pairs of PND bifurcate.  
We remark that if we expand a generic type $I$ Kasner around any point different from the type $D$ one instead a Taylor series always exists.

\section{Teukolsky Master Equation}

Since the Kasner vacuum spacetime with indices $p_1=-1/3$, $p_2=p_3=2/3$
is of Petrov type $D$, we can export the discussion typical of perturbed black hole
spacetimes into the arena of cosmology, applying the Teukolsky
Master equations machinery~\cite{Teukolsky2}. 

To study curvature pertubations here it is convenient to introduce another Newman-Penrose tetrad, obtained from (\ref{tetrade})-(\ref{tetradeb}) 
after a class III null rotation (a boost) which makes $\epsilon$ vanishing. This is
\begin{eqnarray}\label{tetradeK}
l&=&\frac{t^{-p_1}}{\sqrt{2}}\left[ \partial_t +t^{-p_1}\partial_x\right] \,,  \qquad 
n=\frac 1{\sqrt{2}}\left[ t^{p_1}\partial_t-\partial_x\right] \,,\nonumber \\
m&=&\frac 1{\sqrt{2}}\left[ t^{-p_2}\partial_y+it^{-p_3}\partial_z\right] \,, 
\end{eqnarray}
which gives the non zero spin coefficients: 
\begin{eqnarray}
&&\mu =\frac 1{2\sqrt{2}}(1-p_1)t^{p_1-1}=-t^{2p_1}\rho\,,\quad \gamma =-\frac 1{\sqrt{2}%
}p_1t^{p_1-1}\,,\quad 
\nonumber \\
&&\lambda =-\frac 1{2\sqrt{2}}(p_3-p_2)t^{p_1-1}=-t^{2p_1}\sigma\,,\label{spinc}
\end{eqnarray}
and the non zero Weyl scalars: 
\begin{eqnarray}\fl\qquad
&&\psi _0=\frac{t^{-2(p_1+1)}}2p_1(p_2-p_3)=t^{-4p_1}\psi_4\,,  \qquad\psi _2=-\frac 12t^{-2}p_2p_3\,.
\label{PSIs} 
\end{eqnarray}
The advantage of this tetrad is that it results
again principal and canonical ($\psi_0=\psi_4=0$) for the Petrov type $D$ with $p_2=p_3=2/3$; moreover it satisfies
the useful condition on the spin coefficients $\epsilon=0$ (it is a
Kinnersley frame) which allows us to apply {\it in toto} the original Teukolsky
notation. In particular the fact that the tetrad above becomes the canonical one for the Kasner  type $D$ solution with $p_2=p_3=2/3$ ensures the gauge and tetrad invariance of the perturbed $\psi_0$ and $\psi_4$ \cite{Teukolsky2,SW} for the quasi-$D$ models. More in general, any covariantly defined quantity  that vanishes in the background spacetime is a gauge invariant perturbation \cite{SW,BMMS,SB}.

The equations for the gauge and tetrad invariant first order
massless perturbations of various spin-weight $s$ (coincident with the
helicity~\cite{PenroseRindler}) in this background, written for simplicity
in absence of perturbative sources using the frame (\ref{tetradeK})
and its derived NP quantities (\ref{spinc}) and (\ref{PSIs}), are given by
the following NP relations valid for any vacuum type $D$ geometry
\begin{eqnarray}
\fl\quad  &&\{[D-\rho ^{*}+\epsilon ^{*}+\epsilon -2s(\rho +\epsilon
)](\Delta +\mu -2s\gamma )  \label{mia1} \\
\fl\quad  &&-[\delta +\pi ^{*}-\alpha ^{*}+\beta -2s(\tau +\beta )]\,(\delta
^{*}+\pi -2s\alpha )-2(s-1)(s-1/2)\psi _2\}\psi^{(s)} =0  \nonumber
\end{eqnarray}
for spin weights $s=1/2,1,2$ and 
\begin{eqnarray}
\fl\quad  &&\{[\Delta -\gamma ^{*}+\mu ^{*}-\gamma -2s(\gamma +\mu )](D-\rho
-2s\epsilon )  \label{mia2} \\
\fl\quad  &&-[\delta ^{*}-\tau ^{*}+\beta ^{*}-\alpha -2s(\alpha +\pi )](\delta
-\tau -2s\beta )-2(s+1)(s+1/2)\psi _2\}\psi^{(s)} =0  \nonumber
\end{eqnarray}
for $s=-1/2,-1,-2$. The case $s=\pm 3/2$ is available in the literature in
Geroch-Held-Penrose form~\cite{Guven,RARITA2} and finally the case $s=0$ is
given by~\cite{Detweiler} 
\begin{eqnarray}
\fl\quad  &&[D\Delta +\Delta D-\delta ^{*}\delta -\delta \delta ^{*}+(-\gamma
-\gamma ^{*}+\mu +\mu ^{*})D+(\epsilon +\epsilon ^{*}-\rho ^{*}-\rho
)\Delta   \nonumber \\
{}\fl\quad &&+(-\beta ^{*}-\pi +\alpha +\tau ^{*})\delta +(-\pi ^{*}+\tau -\beta
+\alpha ^{*})\delta ^{*}]\psi^{(0)} =0\,.  \label{KGORDY}
\end{eqnarray}
Perturbative sources can be easily added following Teukolsky. Introducing a
``connection vector'' that for our Kasner models has the components 
\begin{equation}
\Gamma ^t=\frac 13t^{-1}\,,\quad \Gamma ^x=-t^{-\frac 23}\,,\quad \Gamma
^y=\Gamma ^z=0\ ,\quad   \label{eq:SPINNOL2}
\end{equation}
such that $\nabla ^a\Gamma _a=0$ and $\Gamma ^a\Gamma _a=4\psi _2$,
all these equations collapse to the unique PDE form~\cite{bcjr,bcjm,bcjTN}: 
\begin{equation}
\lbrack (\nabla ^a+s\Gamma ^a)(\nabla _a+s\Gamma _a)-4s^2\psi _2^{(0)}]\psi
^{(s)}=0\,  \label{eq:bellak}
\end{equation}
where $\psi _2^{(0)}$ is the background Weyl scalar in (\ref{PSIs}).  
Equation~(\ref{eq:bellak}) gives a common structure for these
massless fields in the LRS Kasner background varying the spin index $s$. The
components of the TME with negative spin must be multiplied by a certain
prefactor in order to give the physical components of the fields, as well
explained by Teukolsky \cite{Teukolsky2}. For the aim of our subsequent analysis we shall be
interested in the gravitational case only: consequently one has to consider the
solution $\psi^{(s)} $ of the TME in the case $s=-2$; the Weyl scalar with
negative helicity is then given by $\psi _4=\rho ^4\psi ^{(-2)}$. We recall  that
this approach is possible for the special type $D$ (and type $O$) spacetimes only, because it is this speciality type that 
allows to decouple and separate the perturbative equations in an invariant way
\cite{SW}. In the literature there are mathematical studies of the
perturbations of type $D$ spacetimes in terms of Hertzian and Debye potentials
\cite{Cohen1,Pons,dhur1,dhur2}. Here we prefer to attack the problem 
by using the standard Teukolsky theory instead. The TME admits
separable solutions of the form 
\begin{equation}
\psi ^{(s)}(t,x,y,z)=e^{ik_xx}e^{ik_yy}e^{ik_zz}Y(t)\ ,
\end{equation}
with $Y(t)$ satisfying the ``Master'' equation 
\begin{equation}
\fl\quad t^{-(1+\frac 23s)}\frac d{dt}\left[ t^{(1+\frac 23s)}\frac
d{dt}Y(t)\right] +\left[ k_{\perp }^2t^{-\frac 43}-2isk_xt^{-\frac
23}+k_x^2t^{\frac 23}\right] Y(t)=0,  \label{TEMPOR}
\end{equation}
where $k_{\perp }^2=k_y^2+k_z^2$. We point out that this ODE can be reduced
to a Heun biconfluent hypergeometric equation. In fact, introducing the
rescaling 
\begin{equation}
Y(t)=t^{2/3s}\,e^{i\frac 34k_xt^{4/3}}Z(t)
\end{equation}
and then changing the variable $t$ as follows 
\begin{equation}
t=\left( \frac 2{3k_x}\right) ^{3/4}\,e^{i\frac{3\pi }8}\,\nu ^{3/2},
\end{equation}
Eq. (\ref{TEMPOR}) takes the canonical form~\cite{Exton,RON}: 
\begin{equation}\fl
\nu \frac{d^2Z}{d\nu ^2}+\left( 1+\alpha -\beta \nu -2\nu ^2\right) \frac{dZ
}{d\nu }+\left\{ \left( \gamma -\alpha -2\right) \nu -\frac 12\left[ \delta
+\left( 1+\alpha \right) \beta \right] \right\} \,Z=0\,,
\end{equation}
with 
\begin{equation}
\fl
\alpha =-s\,,\qquad \beta =0\,,\qquad \gamma =3s\,,\qquad \delta =-\frac
92\left( \frac 2{3k_x}\right) ^{1/2}k_{\perp }^2\,e^{i\pi /4}\,.
\end{equation}
Additional perturbative sources can easily be introduced in the problem
once expanded on this complete basis. Or even more simply, our equation
could be easily solved for asymptotic values $t\to 0$ (although in this
limit the curvature invariants become infinite and perturbation theory
clearly has problems) and $t\to \infty $. In analogy with black hole
physics in which, because of the Peeling Theorem, one has that the fields
asymptotically in space are described by radial power laws, here the
physical field components are described by gauge and tetrad invariant
temporal power laws which are dependent on $s$, reminding   ``a
Cosmological Peeling-off Property of Gravity''~\cite{Carmeli}. However in
the following we will not be interested in this more formal
task, but instead we will analyze a very simple subset of perturbations, 
making a link with Section 2. 

We now analyze those special solutions of the TME that represent  Kasner models perturbatively  close to the  type $D$ ($p_1=-1/3$, $p_2=p_3=2/3$) background one. 
 First we write again $p_3=2/3+\varepsilon$ and, taking into account the 
parameter constraints (\ref{constr}), 
expanding the expressions (\ref{PSIs}) we obtain that the (real) exact Weyl scalars become
\begin{equation}
\fl\quad
\psi _0\simeq \frac 13t^{-\frac 43}\varepsilon +o(\varepsilon ^2),\quad \psi
_2\simeq -\frac 29t^{-2}+o(\varepsilon ^2)\,,\quad \psi _4\simeq \frac
13t^{-\frac 83}\varepsilon +o(\varepsilon ^2)\,\,\,  \label{LINEARIZED}
\end{equation}
showing again that despite the PNDs are not analytic on the type $D$ point, the
curvature is. 
 
 Going back to our Master equation (\ref{TEMPOR}), the substitution of $k_x=k_y=k_z=0$ in~(\ref{TEMPOR}) gives
the general solution $Y=c_1+c_2t^{-\frac 23s}$, which for $s=\pm 2$ yields exactly
the linearized Weyl scalars~(\ref{LINEARIZED}) above. Not surprisingly, to recover a Kasner model from the general perturbation equation (\ref{TEMPOR}) we have to assume homogeneity, represented by vanishing wave-numbers ({\it i.e.} infinite wavelengths). 
 
  We point out that for perturbed  black holes
 $\psi _0$ and $\psi _4$ describe gravitational radiation at infinity \cite{Teukolsky2}. 
More in general, using the interpretation provided by Szekeres's 
gravitational compass~\cite{Szekeres},  $\psi _0$ and $\psi _4$ are responsible
for  the transverse deformations with respect to the congruences $l$
and $n$,  including in particular the
possible deformations of gravitational wave type. For the case of our special perturbations the wave vector is zero 
($\vec{k}=0$), implying the absence of propagation of gravitational signals for perturbations that have to represent 
 Kasner solutions close to the LRS type $D$ solution. 
  The absence of gravitational waves in Kasner models is well known, and had to be expected here from the vanishing of the magnetic Weyl tensor (\ref{MW}) and from the manifest spatial homogeneity of the metric. Although  the metric tensor is not a gauge invariant object, 
  the spatial independence of (\ref{LLLL}) directly reflects the invariant spatial  Killing symmetries embodied in  the spatial tetrad vectors (\ref{orto}). However, the Weyl
scalars $\psi _0^{(1)}$ and $\psi _4^{(1)}$ 
that we have derived from the Master equation (\ref{TEMPOR}) are gauge and
tetrad invariant perturbations of the type $D$ background, so that the absence of spatial propagation  is a physical information derived here in a novel way, using the formalism of the Teukolsky equation for gravitational perturbations. 
   
Finally, going back to Szekeres's  gravitational compass~\cite{Szekeres}, the electric Weyl tensor (\ref{EW}) represents the only direct curvature contribution to the Jacobi (or in particular the geodesic deviation) equation, and for any Petrov type $I$ field and any transverse frame  can be re-expressed as in equation (\ref{bella}).  It is actually this expression for $E(u)_{ab}$ that justifies in general (and not just in a perturbative context) the ``transverse frame" terminology \cite{beetle,Szekeres}: for a generic tetrad with $\psi_1\neq0$ and/or $\psi_3\ne0$ there would also be longitudinal contributions to (\ref{bella}) \cite{Szekeres}, and  the magnetic Weyl tensor $H(u)_{ab}$ would be non zero, clearly a ``tetrad gauge" effect. 
For type $D$ spacetimes, observers using a canonical tetrad (and associated orthonormal one) don't measure any transverse contribution.
For the general Kasner case, either considering the preferential observer and associated tetrad (\ref{orto}) with the corresponding null tetrad (\ref{tetrade})-(\ref{tetradeb}), or any other  null tetrad like (\ref{tetradeK}), 
one  has  purely time dependent monotonically decaying
Weyl scalars. Thus, the absence of gravitational waves is once again manifest.  On the other hand, the transverse contributions $\psi_0$ and $\psi_4$ to $E(u)_{ab}$ in (\ref{bella}) are a good example that the presence  of these transverse terms does not necessarily imply the presence of gravitational waves. An analogous results can be obtained in studying the spacetime of stationary axi-symmetric rotating neutron stars \cite{BWMB}.

\section{Conclusions}

In this paper we have considered quasi-$D$ Kasner models that can be parametrized  as linear perturbations of the special  Petrov  type $D$ case.
The simple calculations in Section 2 allow us to conclude  that 
in general a perturbed spacetime will not retain the speciality character of the background and therefore will be of general Petrov type $I$.
The fact that the indicators of speciality like $\mathcal{S}$ or $\mathcal{S}_{\rm norm}$ (which is more significant in the perturbative context)  
may or may not be linear in the perturbations  only reflects the way they are defined. Thus the fact that $\mathcal{S}$ only changes at second order 
 does not necessary imply Petrov speciality for the linearly perturbed spacetimes. In fact we have obtained, as a concrete example, four distinct  PNDs for the quasi-$D$ Kasner solutions. 
In other words, in the case of the SI $\mathcal{S}$, we have to consider 
it as a  quantity in which the first non vanishing perturbative term has to be retained, to be consistent with first order solutions of Einstein equations. This is not too surprising, as $\mathcal{S}$ is not a dynamical variable, but instead by construction is higher order, as the curvature invariants $I$ and $J$. 
It is the solution of Einstein equations that has to admit a Taylor expansion around the given background\cite{waldbook,SW,BMMS,SB}, and it is in solving Einstein equations at first order that higher order terms have to be neglected for consistency. More precisely,  the existence of a parametrised family of solutions admitting a Taylor expansion around a given background is a prerequisite for the application of perturbation methods to Einstein equations\cite{waldbook,SW,BMMS,SB}.
 Once the perturbative solution has been found, this does not imply that {\it any} quantity of interest must be necessarily computed retaining first order terms only.  If the first perturbative terms in such a quantity are of higher order, the first non vanishing perturbative order must be retained.
 Also,  if a Taylor expansion of a spacetime family exists around a given background, this does not imply that any quantity computed for the perturbed spacetime is a Taylor expansion: as shown by our expansion of the PNDs, they do not admit  a Taylor series around the type $D$ background case. 

In this article we have also adopted the Teukolsky Master equation formalism, which is commonly used in black holes perturbation theory to study gravitational radiation.  Importing  this tool in cosmology we have used it here to demonstrate  that a quasi-$D$ Kasner spacetime  does not contain gravitational waves. 
This confirms,  with  a novel approach, the well known absence of gravitational radiation in general type $I$ Kasner spacetimes, shown here by the vanishing of the  magnetic Weyl tensor $H(u)_{ab}$ and also resulting  by the vanishing in Kasner models of the gravitational superenergy flux \cite{BV}.

Finally, we have used this Kasner example to argue  that the presence of transverse curvature contributions in the Jacobi or geodesic deviation equation (represented by the $\psi_0$ and $\psi_4$ terms in the Electric Weyl tensor $E(u)_{ab}$) does not necessarily imply the presence of gravitational radiation. Clarification of the interpretation of different transverse frames, to the end of  choosing those that only contain the gravitational waves degrees of freedom in the transverse part of the curvature at large distances from the source, is essential to the problem of wave extraction from numerical simulations \cite{NBBBP}.

\section*{Acknowledgments}
We are grateful to Bruce Bassett, Emanuele Berti, Jerry Griffiths, Robert
Jantzen, and Giovanni Montani for useful discussions on
the subject.

\end{document}